# ENERGY BALANCING ALGORITHM FOR WIRELESS SENSOR NETWORK

**GHASSAN SAMARA¹, MOHAMMAD A. HASSAN², MUNIR AL-OKOUR³**

1,2,3 Computer Science Department, Zarqa University, Zarqa- Jordan
E-mail: ¹gsamarah@zu.edu.jo, ²mohdzita@zu.edu.jo, ³munir.okour@outlook.com

**ABSTRACT**

A Wireless Sensor Network (WSN) is made up of a large number of nodes that are spread randomly or on a regular basis to detect the surrounding environment and transfer data to a base station (BS) over the Internet to the user. It is widely used in a variety of civil and military concerns. Because the sensor has limited battery capacity, energy efficiency is a critical issue with WSNs. As a result, developing a routing protocol that decreases energy consumption in sensor nodes to extend the lifetime of the WSN using an intelligence algorithm has become difficult. LEACH is the first hierarchical routing protocol that divides the WSN into clusters to reduce energy usage. However, it has reached its limit in selecting a suitable cluster head and the sensor nodes to be joined, as well as their quantity. Thus, this research proposes an algorithm called Wireless Energy Balancing algorithm (WEB) that works on energy balancing distribution by identifying a suitable cluster head with minimum distance and high energy. Then it uses the knapsack-problem as a novel algorithm to design the cluster members. The simulation results demonstrate that the WEB algorithm outperforms LEACH by 31% in terms of energy conservation and WSN lifetime extension.

## 1. INTRODUCTION

A WSN differs from a traditional wired network. It has many nodes and an immense amount of aggregate data; node batteries are difficult to recharge or replace; computational power and storage are restricted; and dynamic network topology. (5), (6), (7), (8), and (9) (8),

Because of their small size, sensor nodes in the WSN have limited resources such as energy, communications, computing, and memory. One of the most essential factors that a researcher must address while establishing a protocol is energy [9, 10, 11 12]. Because a limited battery capacity powers the sensor node, and energy consumption occurs in sensing, data processing, and communications, energy efficiency is a significant challenge in the WSN. Communications is the WSN's largest energy consumer [13, 14, and 15]. As a result, WSN protocols must be energy efficient in order to reduce energy consumption in sensor nodes and increase network lifetime [16, 17].

Hierarchical routing entails categorizing sensor nodes into hierarchical clusters. Each cluster has many sensor nodes with a single head, with the goal of reducing sensor node energy consumption and extend the network's lifetime. LEACH [18, 19] was one of the first hierarchical routing protocols.

The LEACH protocol is the first hierarchical protocol designed to extend the life of a WSN. This protocol ensures that energy is distributed evenly across sensor nodes by grouping nodes into clusters and rotating the cluster head between nodes on a regular basis. The cluster heads collect data received from the sensor node, aggregate it, and transfer it to the BS, resulting in less energy consumed for data transmission to the BS [20].

Despite the fact that LEACH distributes energy on sensor nodes to create a balanced energy network, it does not place constraints on the candidate cluster head. LEACH selects a candidate cluster head at random [21]. If the candidate cluster head is located distant from the BS, it will expend significant energy forwarding data [22, 23, 24, 25]. LEACH does not take into account the candidate cluster head's remaining energy. As a result, a candidate cluster head with low residual energy can be chosen as a Cluster Head (CH), whereas this study treats this by selecting the nearest node to BS with high residual energy; LEACH does not impose any restrictions on the sensor nodes that will be joined with this candidate cluster head or their count, whereas this study innovated a new algorithm by using the knapsack-problem to choose the sensor nodes that will join with their cluster head [26].

Clustering in WSN is an NP-hard problem, a class of decision issues that are as difficult as any problem in NP. No known polynomial algorithm can tell whether a solution is optimal or not given a solution [27].

The Knapsack problem is NP-hard. Each item in a set has a weight and a value, which are used to determine the count of each item to be included in a collection. That is, the total weight is less than or equal to a specified limit, and the total value is as great as possible. If all weights are non-negative integers, a polynomial approach can be used to solve it using dynamic programming [27].

This research is based on the 0/1 Knapsack-problem, in which the cluster is a knapsack and the items are sensor nodes. This inspiration will generate a weight for the cluster as Knapsack, where the item weight is energy and the value is distance. This study will pick the candidate sensor node that will be a cluster head by selecting the closest node to BS with high energy. Then, the cluster members will be created using the dynamic programming approach to solve the 0/1 knapsack problem. The WEB algorithm will extend the lifetime of the WSN by employing this strategy.

The limited energy of the sensor is dependent on the battery, which is difficult to repair or recharge, making energy a critical concern with the WSN. As a result, there are numerous designs and protocols in use to reduce energy usage on sensor nodes. LEACH is the initial protocol of the hierarchical routing protocol, which separates the WSN into clusters. Each cluster consists of a group of sensor nodes with a single head. Sensor nodes monitor their surroundings and relay sensitive data to the cluster head. The data aggregation can eliminate a large amount of duplicated data, reducing the communication load on the CH node, and only the CH can directly transfer aggregative data to the BS. Because it has to process more work than other nodes, the CH node's energy is quickly depleted. Clustering reduces message volume and limits direct communication between sensor nodes and the BS. Despite the benefits of the LEACH protocol in terms of lowering energy consumption and increasing network lifetime, it has certain shortcomings in terms of identifying eligible candidate cluster heads. Being chosen at random may result in an uneven distribution of the overall network and incorrect determination of the suitable nodes or their number that will be joined with their candidate cluster head, causing the sensor node to consume more energy and reduce the WSN lifetime.

## 2. RELATED WORK

Many protocols are used to reduce sensor node energy consumption and work on a restricted power battery that is difficult to replenish or replace. These protocols have been divided into two categories: network structure-based protocols and operation-based protocols. This research is based on the first category of cluster-based routing protocols, which are distinguished by the segmentation of the network into clusters, each of which has a set of nodes known as a cluster. This divide has resulted in a more extended network lifetime.

Authors in [28] created a first protocol, LEACH, that disjointed the network into many clusters. Each cluster has a cluster head that collects data from joined nodes and sends it to the BS. This method has various advantages during routing, such as scalability, reduced energy usage by aggregating data, and reduced transmissions to the BS, which led to the adoption of LEACH in the hierarchical routing protocol.

Power-Efficient Gathering in Sensor Information Systems (PEGASIS) was proposed by [29]. This is a chain-based protocol that improves LEACH algorithms. As a result, PEGASIS is a LEACH protocol enhancement. PEGASIS avoids cluster formation by sending and receiving each node from a neighbor and selecting only one node from that chain to transmit to the base station.

Authors in [30] created the Threshold sensitive Energy Efficient sensor Network protocol (TEEN), which varies from LEACH in that it utilizes a hierarchical method based on clusters of closer nodes. This process continues to the second level till the BS is attained.

Many academics have made enhancements to LEACH, which has many flaws, in order to improve its performance, such as:

In [31], authors improved the LEACH Protocol of WSN (VLEACH) vice-Cluster Head, which aims to reduce energy consumption within the wireless network when the cluster head does not have enough energy to transmit data or collect cluster members to the BS, the vice-Cluster Head takes over as the cluster head of that cluster. Their progress demonstrates that VLEACH outperforms the LEACH approach.

Some studies employ intelligence approaches or combine a WSN to create an intelligent application with routing protocol optimization to conserve limited resources in an extensive collection of sensor nodes and reduce energy consumption in the WSN.

In [32], authors used the PSO algorithm at the BS to create energy-aware clustering for wireless sensor networks. All nodes submit information about their current energy status and locations to the BS for each round, which begins with a setup phase to build a cluster. Following that, the base station uses the PSO algorithm to determine the best number of cluster heads that can minimize the cost function; they present a new cost function to produce better network partitioning with the least intra-cluster distance and cluster heads that are optimally distributed throughout the network.

Authors in [33] improved a LEACH strategy by using the PSO algorithm to enhance and optimize the clustering process by taking into account energy, communication costs, load balance, and other factors to determine the cluster-head node; and to solve LEACH problems such as disparate clustering, the disparate load of the cluster-head node, and other disadvantages to effectively balance and prolong the WSN lifetime.

Authors in [34] proposed KPSO, a new hybrid clustering algorithm based on K-Means clustering and Particle Swarm Optimization (PSO), to overcome the energy consumption problem in WSNs based on clustering. The K-means algorithm was used to split the network into a preset number of clusters based on the distance between an elected cluster head and the remainder of the nodes in the same cluster. The PSO algorithm was used to search for the optimal CH within each cluster created by the K-means. The algorithm (KPSO) was examined and compared with the LEACH protocol and K-Means clustering. It demonstrated an average improvement of around 49 percent over the LEACH protocol and approximately 18 percent over the K-Means clustering algorithm.

In [35] authors suggested a new method based on the genetic algorithm GA to determine the optimal number of CHs and the agent cluster head to ensure that CH data reaches the BS. This strategy has increased the residual energy of each node. When compared to LEACH, the results revealed a reduction in energy usage and increased network longevity.

## 3. THE WIRELESS SENSOR NETWORK ENERGY BALANCIng (WEB)

The key prerequisite before modelling development is the collection of data for a certain process. A sensor node is a component of a more extensive network of sensor nodes. Each sensor node in the WSN must collect data from its surroundings and send it to cluster heads before sending it to the power BS.

The WEB algorithm that will be developed includes multiple steps to achieve the ultimate goal of reducing sensor node energy consumption and extending the lifetime of the WSN. These stages are thoroughly discussed below:

Step 1: Data collection via base station after each node transmits to the BS information about its current position (perhaps determined by employing a GPS receiver) and residual energy level.

Step 2: Using the Euclidean distance formula, calculate the distance between each sensor node and the BS.

Step 3: Select a prospective cluster head with a short distance and high energy to become the cluster head.

Step 4: Using the Euclidean distance formula, calculate the distance between the cluster head and each sensor node.

Step 5: Determining the lost energy of the cluster, which is the energy consumption in the cluster head to transmit or receive one bit, adding the energy consumption in the cluster head of aggregated data, and then dividing the outputs by the energy consumption in a cluster member, which is represented in the energy consumption of the cluster member to transmit one bit.

Step 6: Calculate the distance weight, the distance between the cluster head and the BS multiplied by the length of a field and divided by the result of multiplying the field's dimensions.

Step 7: Calculate the cluster weight by reflecting the lost energy of the cluster in step five, adding the weight of the distance in step six, and multiplying by the reset of alive nodes.

Step 8: Create a cluster using the knapsack-problem algorithm, which connects cluster members to their cluster leader.

Step 9: If there are still nodes, go back to step 1.

Step 10: Calculate the energy usage of each node.

Step 11: Exit the algorithm if all nodes have died. Otherwise, a new round will be started from step 1.

The flowchart of these processes is depicted in Figure 1:

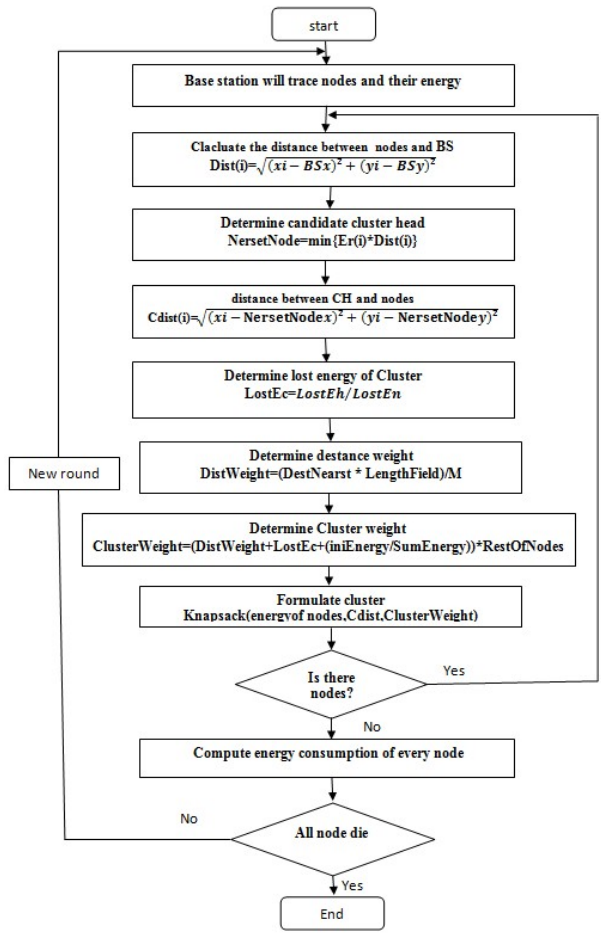

*Figure 1. Flowchart of WEB algorithms*

Construct WEB Algorithm

To reduce sensor node energy consumption and increase WSN lifetime, the protocol must consider the factors that influence energy consumption in the sensor node. Energy consumption is influenced by communication procedures, data aggregation, the distance between sensor nodes, the distance between the head and the BS, and data processing. According to Equation 1, one of the most essential things that consume energy is distance.

The WEB algorithm will be separated into rounds, with each round consisting of two phases: LEACH algorithm operation, setup phase, and steady-state phase.

During the setup phase, the WEB algorithm will identify the candidate cluster head, which member nodes will be linked to their cluster, and the time distributed on each node to provide its sensed data, as detailed in the steps below.

(i) Setup phase

Step 1 Cluster-head selection: The first step Cluster-head choice: Each network node sends its location and energy information to the BS.

To choose a cluster head, BS will perform the following calculation:

Using the Euclidian Formula, calculate the distance between BS and each node, as indicated in Equation 1.

$$Dist(i) = \sqrt{(BSx - Xi)^2 + (BSy - Yi)^2} \quad (1)$$

Where (BSx, BSy) is the coordinates of BS and (Xi, Yi) is the coordinates of node i.

Equation 3.2 demonstrates how to select the candidate cluster head.

$$NersetNode = min\{Dist(i) \times Ei\} \quad (2)$$

Where NersetNode is the location of a candidate cluster node, $E_i$ is the energy of node i and Dist(i) is the distance between node i and BS. For the first objective, reducing energy consumption in the cluster head, this study uses this Equation to select a cluster head that is close to BS, which means that the distance is less to consume energy. After a few rounds as the cluster head, the CH node passes to another node with high energy.

Step 2 Cluster formation: Clusters are formed using the following formula:
, which is the distance between candidate cluster head and BS.

$$distNtoBS = Dist(NersetNode) \quad (3)$$

is the cluster head selected from 2. The WEB algorithm will calculate Cdist(i) the distance between each node and the candidate cluster head using Euclidian Formula as shown in Equation 4.

$$Cdist(i) = \sqrt{(xi - NersetNodex)^2 + (yi - NersetNodey)^2} \quad (4)$$

To find the maximum distance in the network area, use Euclidian Formula as in Equation 5.

$$Dmax = \sqrt{Length^2 + Width^2} \quad (5)$$

Where Dmax is the largest distance in the field and (Length, Width) is a constant variable as maintained in table 1.

This study will use the above Equations to determine the weight of each factor and its effect on energy consumption.

1- Distance

The most critical element in energy usage is distance. When the distance between two points increases, so does the amount of energy consumed. So, Equation 6 was employed in this study to determine an appropriate weight to reduce energy usage.

$$DistWeight = (dsetHtoBS * Length)/Dmax^2 \quad (6)$$

2- Communication and data aggregation processes

In addition to data aggregation, communication procedures include energy usage in transmitting or receiving packets at a non-head node or a head node.

In addition to the data aggregation process indicated in Equation 7, a minimal amount of energy is lost in a head node when transmitting a packet to BS.

$$lostEnergyH = (Elec + EDA) * PacketLength + Elec * ctrPacketLength \quad (7)$$

When a transmission packet is sent to the cluster head, the non-head node loses the least amount of energy, as shown in Equation 8.

$$lostEenergyN = Elec * ctrPacketLength \quad (8)$$

Equation 9 will be used to identify a suitable weight to decrease communication operations and data aggregation.

$$LostEnergyC = LostEnergyH / LostEnergyN \quad (9)$$

3- Balance

Equation 10 will be used to determine the number of members cluster for each cluster that ensures a fair distribution of Nodes within the WSN.

$$BalanceWeight = \frac{iniEnergy}{SumOfEnergy} * b \quad (10)$$

Where $iniEnergy$ is a constant variable that contains the initial power of sensor nodes, $SumOfEnergy$ is a variable containing the total energy residual in live sensor nodes, and b is the number of a live sensor node.

As a result, Equation 11 will employ weight for the cluster, which comprises distance weight, lost communication weight, and network balancing weight.

$$ClusterWeight = DistWeight + LostEnergyC + BalanceWeight \quad (11)$$

Finally, in order to determine which node will be linked to which cluster, this study developed a new algorithm that uses the Knapsack-algorithm to attach the node to its cluster head by executing the KnapSack (v,w,n, W). The cluster is formulated as shown in Figure 5, where v is the number of energy nodes, w is the distance between the cluster head and the nodes, which is calculated in Equitation 4, n is the number of residual nodes (alive), and W is the weight of the cluster, which is calculated in Equitation 11.

Step 3: Schedule TDMA and CDMA: The cluster head creates TDMA time slots and distributes them to each member of its cluster and chooses a CDMA code to send sensed data to the BS.

The setup phase procedures will be repeated from step 1 to step 3 until all nodes are connected to their cluster.

(ii) The steady-state phase

After creating all clusters in the WSN, the time slot assigned to each node will run utilizing TDMA and CDMA for each cluster head to transfer the aggregated data to the BS. The BS then computes each node's energy usage to begin a new round until all nodes die. The flowchart of the WEB algorithm is shown in Figure 1.

After the WEB algorithm is completed, it will be run on simulation software, followed by the LEACH algorithm, to study and analyze their outputs for comparison and prediction of results.

## 4. SIMULATION AND RESULTS

BS will collect sensor node data and energy and run the WEB algorithm on MATLAB r2010a with some reasonable network model assumptions such as all sensor nodes deployed within a square area with one BS on the centric; homogeneous, static nodes, and static-sink network; all sensor nodes and the BS are fixed after deployment. All sensor nodes rely solely on the initial battery power, which is non-rechargeable. GPS or a similar positioning method is used to locate all sensor nodes.

The parameters used in MATLAB r2010a for algorithms are listed in Table 1:

*Table 1: Parameter settings*

| Parameter | Description | Initial Value |
|---|---|---|
| NumNodes | Number of sensor nodes | 100 |
| Length, Width | Length and Width of the yard | 100, 100 |
| sinkX, sinkY | X and Y coordination of BS | 50, 50 |
| iniEnergy | Initial energy of each node | 0.5J |
| Eelc | Energy for transferring or receiving of each bit | 50 nJ/bit |
| transEnergy | Energy for transferring of each bit | 50 nJ/bit |
| recEnergy | Energy for receiving of each bit | 50 nJ/bit |
| fsEnergy | Energy of free space model | 100 pJ/bit/m$^2$ |
| mpEnergy | Energy of multi path model | 0.013 pJ/bit/m$^4$ |
| aggrEnergy | Energy of data aggregation | 5 nJ/bit |
| packetLength | Packet size from head to BS | 6400 bit |
| crtpacketLength | Packet size from node to cluster head | 200 bit |

The two-part scenario consists of a scenario action for the WEB algorithm and a scenario action in the same input entered into the LEACH algorithm to compare the results and the scenario of the WEB algorithm as follows: Building a network of 100 sensor nodes and their parameter settings as shown in Table 1 and Figure 2 that shows the distributed sensor nodes and BS where the blue dot stands for the sensor nodes and the green dot stands for the BS at the center position area.

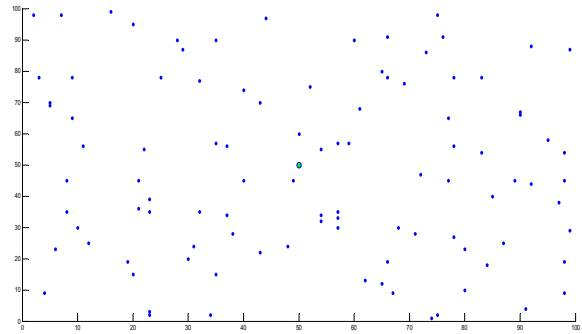

*Figure 2. The distributed sensor nodes*

1- The WEB algorithm is made up of rounds, with each round doing the following tasks:
　i) Each WSN node transmits its position and residual energy.
　ii) In the BS, the distance between sensor nodes is calculated, and the BS then identifies the cluster head and calculates the weight cluster.
　iii) By solving the Knapsack-Problem with dynamic programming to identify which node will link with its cluster leader.
　iv) Following that, it creates all clusters and computes the energy usage of each node and head.
2- A fresh round will be started until all nodes have died.
3- In this instance, all results will be stored in a matrix.
The LEACH algorithm scenario will do the following:
1- Create the WSN as a WEB strategy.
2- The LEACH algorithm is made up of rounds, with each round doing the following tasks:
　i) After all nodes have sent their position and residual energy, use Equation 2.1 to calculate the threshold for randomly selecting cluster heads.
　ii) Forms a cluster as each node connects to its cluster head, which has a strong signal.
　iii) Determines each node's energy consumption.
3- A new round will be started until all nodes have died.
4- Save all results in the matrix.
Simulation Analysis and results
This study shall evaluate and compare the data to determine whether the algorithm performs better in terms of energy consumption in the sensor node and network lifetime extension.
First, Figure 3 demonstrates how the WEB algorithm selects cluster heads when performed on MATLAB R2010a for one round only. (+) red plus denotes cluster head, and (*) green star means nodes joining to their cluster number.

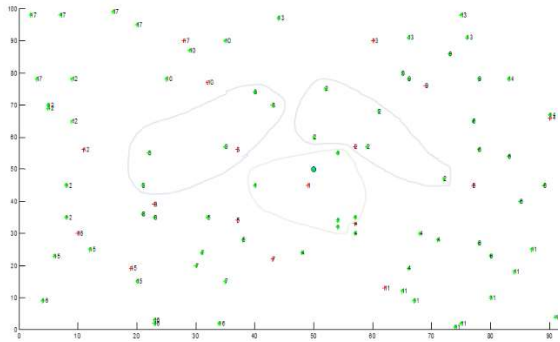

*Figure 3. WEB algorithm formulates clusters for one round.*

Unlike the LEACH algorithm, the WEB algorithm elects the first cluster head that is closest to BS because the energy in this round is initialized with the same energy, then joins the nodes that are closest to their cluster head, and so on, without determining how many cluster heads must be available. As a result, the number of cluster heads varies from round to round, depending on the weight of each cluster. Furthermore, the node distribution is better since the distance between each node and its cluster head is close, resulting in balanced distributed energy on all WSN and sending more packets with less energy consumption at the sensor node for the entire surrounding area. WSN becomes more dependable and adaptable in this manner.

While the LEACH method chooses the cluster head at random, as illustrated in Figure 4, no more than 0.05 percent of all nodes are chosen on each round [28], and the nodes that join with its cluster are far away.

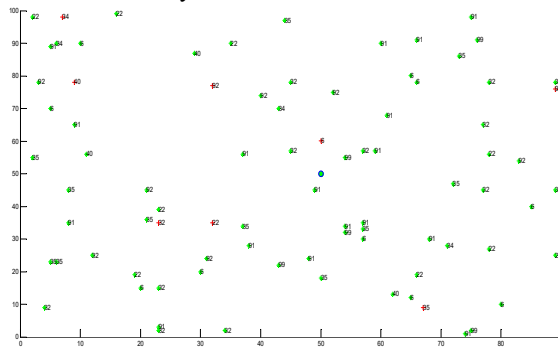

*Figure 4. LEACH algorithm formulates clusters for one round.*

Figure 5 depicts the dispersed cluster heads and their join nodes after two hundred rounds of the WEB algorithm. It should be noted that cluster heads alter depending on their energy and distance. Furthermore, the number of linked nodes varies according to their weight, as defined by Equation 9. In general, the distributed energy style is balanced, as demonstrated in Figures 3 and 5; however, the LEACH algorithm is not, as shown in Figure 6.

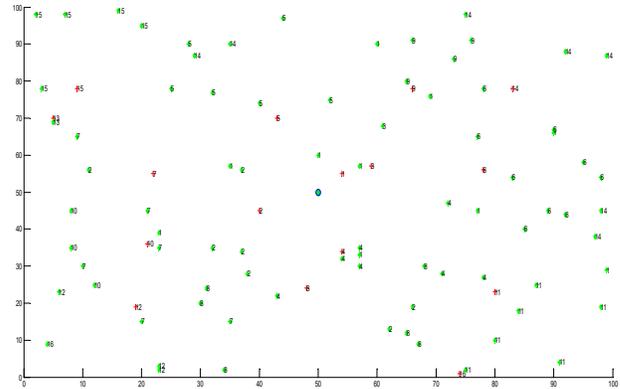

*Figure 5. WEB algorithm formulates clusters after two hundred.*

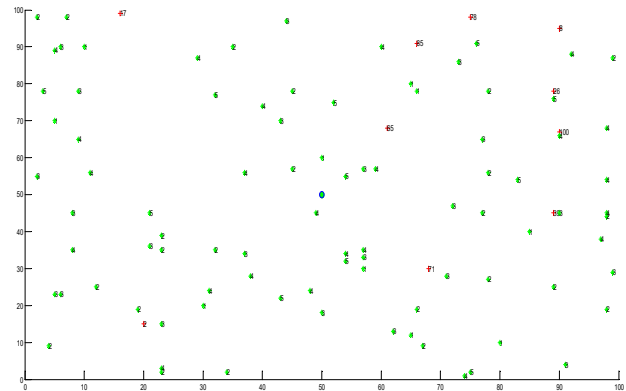

*Figure 6. LEACH algorithm formulates clusters after two hundred.*

Figure 7 depicts dead nodes after seven hundred rounds of algorithm execution. The first dead node in the LEACH algorithm is 76 at round 191, while the first dead node in the WEB algorithm is 32 at round 596, indicating that the WEB algorithm's style is stable and has approximately 3 times better performance for the first node to die rating 31 percent saving consuming energy than the LEACH algorithm.

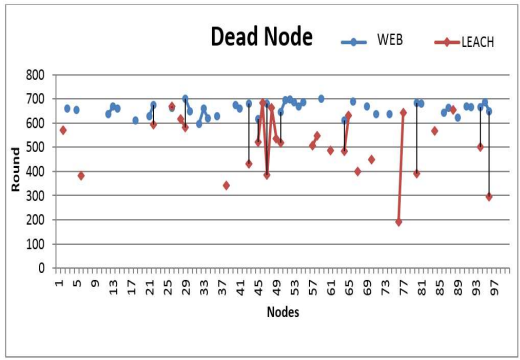

*Figure 7. Node died after algorithms executed seven hundred rounds.*

Figure 8 depicts residual energy in a WSN after seven hundred rounds of the WEB algorithm, demonstrating that the WEB algorithm outperforms the LEACH method of dead nodes (43 in the LEACH algorithm vs 28 in the WEB algorithm). This means that when the method is implemented for one-third of the time, the number of dead nodes (0.28) and energy usage (0.38) are approximately one-third of the time. This signifies that the WSN is in balance. Table 2 displays the amount of dead nodes and excess energy after seven hundred rounds of algorithm execution.

*Table 2: Comparing Algorithms After Executing 700 Round*

| After algorithms execute 700 rounds | WEB algorithm | LEACH algorithm |
|---|---|---|
| Dead Node | 28 dead nodes | 43 dead nodes |
| Residual Energy | 19.70108 from 50 | 2.644393 from 50 |
| Dead Node number | 76 | 32 |
| Dead Node round | 191 | 596 |

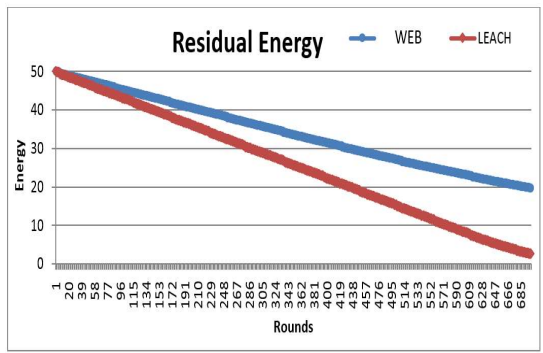

*Figure 8. Residual Energy after the algorithms execute seven hundred rounds*

The data transfer in the WEB method increases slowly after 700 hundred rounds, as shown in Figure 9, and the number of packets at round 700 is 6243. Still, the data transfer in the LEACH algorithm was substantial, with a total of 19679 packets. This means that the demand on the WEB method is approximately three times smaller than the strain on the LEACH algorithm.

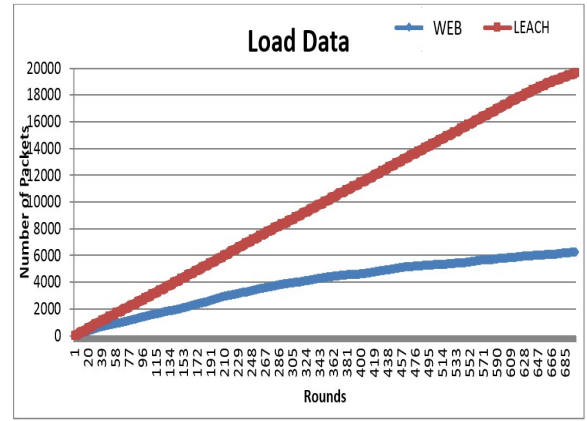

**Figure 9.** Load Data after the algorithms execute 700 rounds.

When the algorithms run until all nodes die, as illustrated in Figure 10, the energy consumption in the LEACH algorithm was discovered to be quickly collapsing until node nine at round 826 and the WSN inertia at round 827 and higher. The LEACH method selects the cluster head at random based on a random integer, assigns it to a node, and tests it against a threshold. Because that random number is bigger than the threshold, no cluster head is elected, as indicated in Equation 1. In contrast, the WEB method finds that energy gradually diminishes until the final node dies, as shown in Figure 11

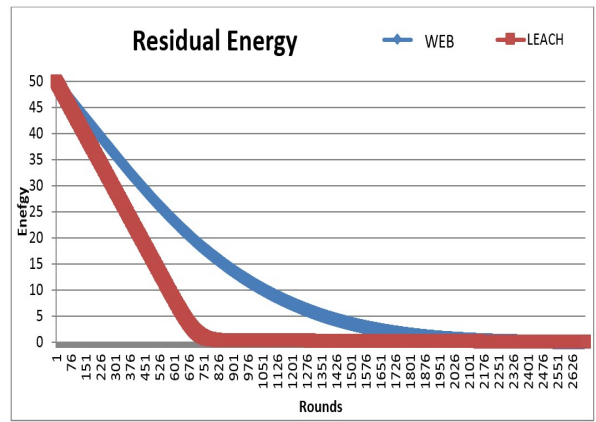

*Figure 10. Residual Energy after the algorithms execute until all nodes die.*

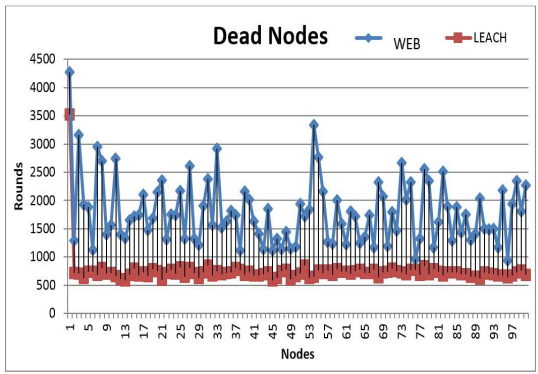

*Figure 11. Dead Node after the algorithms execute until all nodes die.*

*Table 3: comparing algorithms where node ninety-nine dies*

| After algorithms execute | WEB algorithm | | LEACH algorithm | |
|---|---|---|---|---|
| | Node number | round | Node number | round |
| Dead Node | 3 | 2449 | 31 | 849 |
| Residual Energy | 3 | 0.113 365 | 31 | 0.415 072 |

Table 3 shows that when Node 99 dies, the residual energy in the WSN of the WEB algorithm was 0.11 at round 2449, whereas it was 0.41 at round 849 in LEACH. This means that the last node in a WEB algorithm dies after 255 rounds, whereas the last node in LEACH dies after 1851 rounds, implying that the WSN in LEACH is inactive most of the time.

Table 4 compares the WEB method to the LEACH algorithm and shows some discrepancies between the approaches.

Table 3 shows that when Node 99 dies, the residual energy in the WSN of the WEB algorithm was 0.11 at round 2449, whereas it was 0.41 at round 849 in LEACH. This means that the last node in a WEB algorithm dies after 255 rounds, whereas the last node in LEACH dies after 1851 rounds, implying that the WSN in LEACH is inactive most of the time.

Table 4 compares the WEB method to the LEACH algorithm and shows some discrepancies between the approaches.

*Table 4: a comparison between the WEB algorithm and LEACH algorithm*

| WEB algorithm | LEACH algorithm |
|---|---|
| The cluster head identifies regularly. | The cluster head identifies randomly. |
| Each round must contain at least one cluster head. | The round may/may not contain any cluster heads. |
| Determines the cluster head, then formulates cluster. | Firstly determines all cluster heads, then formulates cluster. |
| Maximum numbers of cluster heads are not defined. | Maximum numbers of cluster heads are defined. |
| Number of nodes that are joined to their cluster head vary according to cluster weight. | Number of nodes that are joined to their cluster head vary according to the signal node strength. |

## 5. CONCLUSIONS AND FUTURE WORK

In this study, an appropriate cluster head was chosen based on the nearest node to the BS with high residual energy to ensure a reduction in energy consumption at the cluster head. It distributes the distance between cluster heads over the network in an equitable manner, resulting in a network with balanced energy consumption. Furthermore, combining the cluster members with their nearest cluster heads and the number of cluster members depending on the weight of the cluster leads to lower energy usage for all sensor nodes. Reducing energy consumption in the cluster head and cluster members reduces energy consumption in the overall network, and the WSN's lifetime is extended.

The results demonstrated that the WEB method outperforms the LEACH algorithm every time. The WEB algorithm operates by gradually reducing energy usage, ensuring a healthy CH distribution. By lowering the amount of packets sent between sensor nodes, the WEB algorithm lowered the data burden on the WSN. The WEB algorithm's stability was 31%. As a result of all of this, the WEB algorithm outlasted the LEACH algorithm in terms of WSN lifetime.

In the future, another artificial intelligence algorithm, such as Fuzzy logic or Ant Colony, will be deployed.

REFERENCES

[1] Kumar, A. and Singh, D., 2016. Importance of Energy in Wireless Sensor Networks: A Survey. *An International Journal of Engineering Sciences*, *17*, pp.500-505.

[2] Samara, G. and Aljaidi, M., 2019. Efficient energy, cost reduction, and QoS based routing protocol for wireless sensor networks. *International Journal of Electrical & Computer Engineering (2088-8708)*, *9*(1).

[3] Samara, G., Alsalihy, W.A.A. and Ramadass, S., 2011. Increasing Network Visibility Using Coded Repetition Beacon Piggybacking. *World Applied Sciences Journal*, *13*(1), pp.100-108.

[4] Samara, G. and Blaou, K.M., 2017, May. Wireless sensor networks hierarchical protocols. In *2017 8th International Conference on Information Technology (ICIT)* (pp. 998-1001). IEEE.

[5] Samara, G., Al-Okour, M. 2020. Optimal number of cluster heads in wireless sensors networks based on LEACH, *International Journal of Advanced Trends in Computer Science and Engineering, , 9(1), pp. 891–895*.

[6] Samara, G. and Aljaidi, M., 2018. Aware-routing protocol using best first search algorithm in wireless sensor. *Int. Arab J. Inf. Technol.*, *15*(3A), pp.592-598.

[7] Samara, G., Albesani, G., Alauthman, M., Al Khaldy, M., Energy-efficiency routing algorithms in wireless sensor networks: A survey, *International Journal of Scientific and Technology Research, 2020, 9(1), pp. 4415–4418*.

[8] Samara, G., 2020, November. Wireless Sensor Network MAC Energy-efficiency Protocols: A Survey. In *2020 21st International Arab Conference on Information Technology (ACIT)* (pp. 1-5). IEEE.

[9] Samara, G., Ramadas, S., Al-Salihy, W.A.H. 2010. Safety message power transmission control for vehicular Ad hoc Networks. **Journal of Computer Science**,, 6(10), pp. 1056–1061

[10] Samara, G., Al-Salihy, W.A. and Sures, R., 2010, May. Efficient certificate management in VANET. In *2010 2nd International Conference on Future Computer and Communication* (Vol. 3, pp. V3-750). IEEE.

[11] Samara, G., Abu Salem, A.O. and Alhmiedat, T., 2013. Dynamic Safety Message Power Control in VANET Using PSO. *World of Computer Science & Information Technology Journal*, *3*(10).

[12] M Khatari, G Samara, 2015, Congestion control approach based on effective random early detection and fuzzy logic, *MAGNT Research Report, Vol.3 (8). PP: 180-193*.

[13] Samara, G., Ramadas, S. and Al-Salihy, W.A.. 2010. Design of Simple and Efficient Revocation List Distribution in Urban areas for VANET's, (IJCSIS) International Journal of Computer Science and Information Security, Vol. 8, No. 1.

[14] Samara, G., Alsalihy, W.A.H.A. and Ramadass, S., 2011. Increase emergency message reception in vanet. *Journal of applied sciences*, *11*(14), pp.2606-2612.

[15] Samara, G. and Alsalihy, W.A.A., 2012, June. A new security mechanism for vehicular communication networks. In *Proceedings Title: 2012 International Conference on Cyber Security, Cyber Warfare and Digital Forensic (CyberSec)* (pp. 18-22). IEEE.

[16] Yetgin, H., Cheung, K.T.K., El-Hajjar, M. and Hanzo, L.H., 2017. A survey of network lifetime maximization techniques in wireless sensor networks. *IEEE Communications Surveys & Tutorials*, *19*(2), pp.828-854.

[17] Samara, G., 2018. An intelligent routing protocol in VANET. *International Journal of Ad Hoc and Ubiquitous Computing*, *29*(1-2), pp.77-84.

[18] Samara, G. and Alsalihy, W.A.A., 2012. Message broadcasting protocols in VANET. *Information Technology Journal*, *11*(9), p.1235.

[19] Samara, G., 2020. Intelligent reputation system for safety messages in VANET. *IAES International Journal of Artificial Intelligence*, *9*(3), p.439.

[20] Keswani, K. and Bhaskar, A., 2016. Wireless sensor networks: A survey. *Futuristic Trends in Engineering, Science, Humanities, and Technology FTESHT-16, p.1*.

[21] Samara, G., 2019. An improved CF-MAC protocol for VANET. *International Journal of Electrical & Computer Engineering (2088-8708)*, *9(4)*.

[22] Samara, Ghassan. 2020. Lane prediction optimization in VANET. *Egyptian Informatics Journal*.

[23] Alhmiedat, T. and Samara, G., 2017. A Low Cost ZigBee Sensor Network Architecture for Indoor Air Quality Monitoring. *International Journal of Computer Science and Information Security (IJCSIS)*, *15(1)*.

[24] Salem, A.O.A., Samara, G. and Alhmiedat, T., 2014. Performance Analysis of Dynamic Source Routing Protocol. *Journal of Emerging Trends in Computing and Information Sciences*, *5*(2).

[25] Salem, A.O.A., Alhmiedat, T. and Samara, G., 2013. Cache Discovery Policies of MANET. *World of Computer Science & Information Technology Journal*, *3*(8).

[26] Alhmiedat, T.A., Abutaleb, A. and Samara, G., 2013. A prototype navigation system for guiding blind people indoors using NXT Mindstorms. *International Journal of Online and Biomedical Engineering (iJOE)*, *9*(5), pp.52-58.

[27] Jin, S., Zhou, M. and Wu, A.S., 2003, July. Sensor network optimization using a genetic algorithm. In *Proceedings of the 7th world multiconference on systemics, cybernetics and informatics* (pp. 109-116).

[28] Heinzelman, W.R., Chandrakasan, A. and Balakrishnan, H., 2000, January. Energy-efficient communication protocol for wireless microsensor networks. In *Proceedings of the 33rd annual Hawaii international conference on system sciences* (pp. 10-pp). IEEE.

[29] Lindsey, S. and Raghavendra, C.S., 2002, March. PEGASIS: Power-efficient gathering in sensor information systems. In *Proceedings, IEEE aerospace conference* (Vol. 3, pp. 3-3). IEEE.

[30] Manjeshwar, A. and Agarwal, D.P., 2001, April. TEEN: A routing protocol for enhanced efficiency in WSNs. In *1st International workshop on parallel and distributed computing issues in wireless networks and mobile computing*.

[31] Yassein, M.B., Khamayseh, Y. and Mardini, W., 2009, June. Improvement on LEACH protocol of wireless sensor network (VLEACH. In *Int. J. Digit. Content Technol. Appl. 2009*.

[32] Latiff, N.A., Tsimenidis, C.C. and Sharif, B.S., 2007, September. Energy-aware clustering for wireless sensor networks using particle swarm optimization. In *2007 IEEE 18th international symposium on personal, indoor and mobile radio communications* (pp. 1-5). IEEE.

[33] Xu, Y. and Ji, Y., 2011, September. A clustering algorithm of wireless sensor networks based on PSO. In *International Conference on Artificial Intelligence and Computational Intelligence* (pp. 187-194). Springer, Berlin, Heidelberg.

[34] Solaiman, B., 2016. Energy optimization in wireless sensor networks using a hybrid k-means pso clustering algorithm. *Turkish Journal of Electrical Engineering & Computer Sciences*, *24*(4), pp.2679-2695.

[35] Aziz, L., Raghay, S., Aznaoui, H. and Jamali, A., 2016, March. A new approach based on a genetic algorithm and an agent cluster head to optimize energy in Wireless Sensor Networks. In *2016 international conference on information technology for organizations development (IT4OD)* (pp. 1-5). IEEE.